# Time-encoded single-pixel 3D imaging

Jiajie Teng, Qiang Guo, Minghua Chen, Sigang Yang, Hongwei Chen[*]
Beijing National Research Center for Information Science and Technology (BNRist)
Department of Electronic Engineering, Tsinghua University, Beijing, 100084, China
*Corresponding author: chenhw@tsinghua.edu.cn

**Recently three-dimensional (3D) imaging achieves tremendous success in consumer and industrial manufacturing. However, current 3D imaging methods rarely observe dynamical events happening in a short time interval due to the imaging speed limitation. Here we propose a time-encoded single-pixel 3D (TESP-3D) imaging method based on active infrared imaging with high-speed structured illumination, which can acquire the 3D information of dynamic events at an ultra-high frame rate. This method uses single-pixel photodetectors instead of conventional CCD or CMOS devices as the imaging sensors to overcome the image capturing speed limitation. As a proof of concept, we perform continuous real-time 3D scanning imaging at a record 500,000 fps and acquire 3D information of objects moving at a speed of up to 25 m/s. We anticipate that this technology will have important applications in industrial on-line inspection and high throughput 3D screening.**

In recent years, three-dimensional (3D) imaging technologies[1-3] have gained popularity worldwide, and been applied in various fields including on-line inspection[4], biomechanics[5] and wearable AR/VR devices[6]. The commercially available real-time 3D imaging devices including Microsoft Kinect, Intel Real sense and iPhone X depth camera have already come to people's daily life. On industrial processes, the demand for high-speed measurement, inspection and quality control makes 3D imaging systems essential for non-contact data acquisition[7,8]. Thus, the need for faster 3D imaging technology is becoming urgent in many practical applications as well as theoretical interests.

Currently, there are two main high-speed 3D imaging technologies: the passive and the active methods[9,10]. The stereo vision is the representative technique of passive imaging[11-14], which utilizes pixelated cameras to capture images from at least two different perspectives. The speed limitation comes from the frame rate of the cameras and the complicated imaging reconstruction process, which generally results in a rate below 1000 frames per second (fps)[15]. On the other hand, the active 3D imaging has much better performance due to the artificial illumination. The structured light[16-19], also named as fringe projection profilometry, is widely used for active imaging. It wisely exploits an active projector instead of one of the cameras in stereo vision, and produces coded illumination patterns to recognize each point on the surface. Consequently, the computation time gets decreased while the reconstruction accuracy gets improved. Many structured light approaches have been proposed and investigated for fast 3D imaging[9,20]. Generally, a digital light processing (DLP) device is used to create illumination patterns at 10s kHz refresh rate[9], which also limits the imaging speed besides the imaging sensor's frame rate. So the active 3D imaging method also confronts with the speed bottlenecks from the pattern generation devices and imaging sensors.

In this paper, we propose a novel time-encoded single-pixel 3D imaging technology (TESP-3D) to increase the frame rate significantly. This method breaks through both two speed limitations in active 3D imaging. First, an ultrafast time-encoded structured illumination method is employed, using time-stretch and electro-optical modulation techniques[21-23] to realize 50 MHz or even faster light pattern generation. These patterns are set as the sinusoidal ones to obtain the spatial spectrum of the scenes[24]. Besides, the image sensors are substituted by single-pixel detectors (SPD)[25-27] and located in different spatial angles, which can acquire the optical signals at a much higher rate (above 100 MHz) compared with traditional charge-coupled devices (CCD) or complementary metal oxide semiconductor (CMOS) devices. The data stream captured by each SPD is calculated through inverse fast Fourier transform (IFFT) to reconstruct the original image with different shade information. Then with the photometric stereo algorithm[14], the final 3D reconstruction is easily implemented in a fast speed. As a result, TESP-3D system has the capability for real-time 3D visualization of fast moving objects with a simple and cost-effective system configuration.

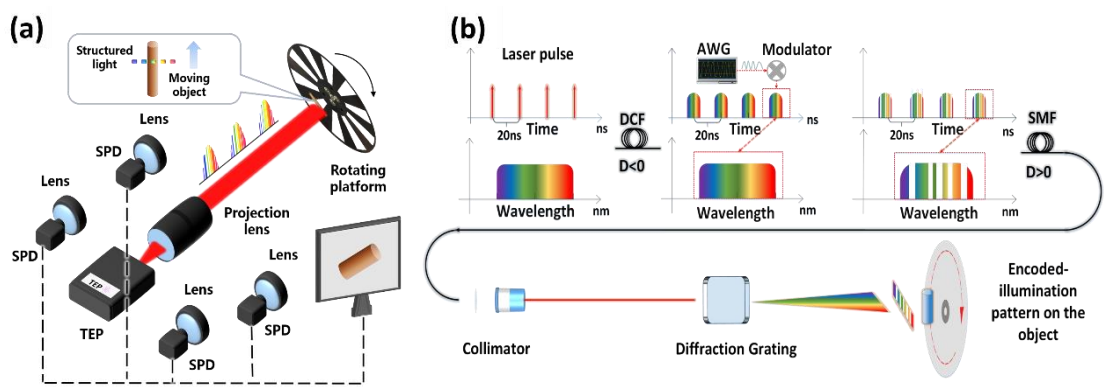

**Figure 1. Schematic of TESP-3D.** (a) System setup: a time-encoded projector (TEP) is utilized for phase-shifting sinusoid structured illumination generation, the structured light passes through the projection lens and forms a line spot on the imaging plane, the object is fixed on a fast rotating chopper, four SPDs located at different positions detecting the diffuse light from the objects, the final 3D image is generated by the reconstruction program in a computer. (b) The principle of TEP: optical pulses are dispersed by the dispersion compensation fiber (DCF), then modulated with sinusoid patterns (generated by an arbitrary waveform generator (AWG)) both in time and wavelength domain; after the single-mode fiber (SMF), the pulses are compressed and fed into space by a collimator; the spatial patterns are generated by the spatial angular dispersion of a diffraction grating.

The system configuration is shown in Fig. 1(a). The phase-shifting sinusoidal light patterns are generated by a time-encoded projector (TEP), whose details are shown in Fig. 1(b), instead of traditional DLP devices. Each time stretched pulse is encoded by sinusoidal signals with varying phases and frequencies through a Mach-Zehnder modulator. The sinusoidal patterns transferring from time to wavelength and space dimension are shown in Fig. 2(a), which proves the feasibility of the time-encoded method. Through successive sinusoidal pattern projection after a diffraction grating, the Fourier spectrum of a scene can be obtained using single-pixel detectors. The pattern generation rate is the same as the pulse repetition rate, which is 50 MHz in our demonstration. The time-encoded pulse train reflected from the scene is shown in the Fig.2(b). Each measurement is finished in 20ns equal to the pulse repetition period. Within each frame, 80 measurements are used for Fourier spectra information acquisition[24,28], and 0.4-μs gap is kept for signal synchronization. The frame duration is 2 μs corresponding to a frame rate of 500,000 fps, which is a record for continuous 3D imaging. Supplementary Movie 1 illustrates the whole imaging process of TESP-3D.

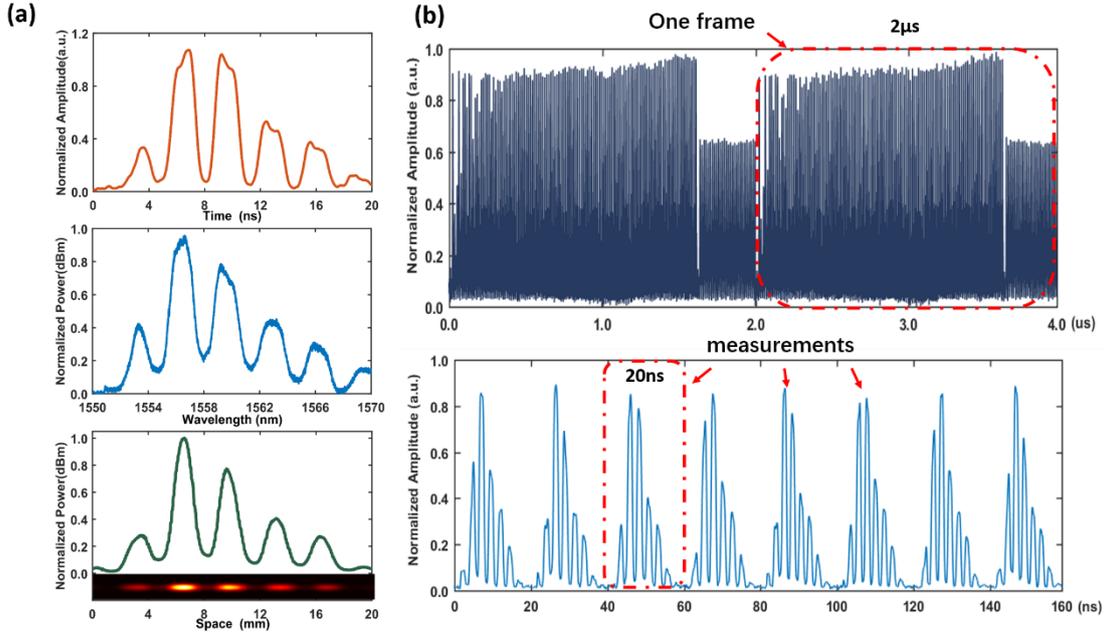

**Figure 2. Basic operations.** (a) The temporal, wavelength, spatial waveforms of one optical pulse, the consistency between them validates the operation of time-encoded projector. (b) The pulse trains after the modulator (below) and captured by one SPD (up) are displayed on the oscilloscope. The measurement period is 20ns same as the pulse period. One frame time is 2μs including 80 measurements and 0.4μs vacancy for the synchronization.

By means of Fourier spectrum single-pixel imaging[11,24,28,29], TESP-3D has the ability of capturing multi-viewing images with different shadow information using four cost-effective SPDs. Providing the inverse detecting vectors exist and the reflectance factor are known and uniform, we can calculate the surface normal in each pixel based on the photometric stereo[14]. Then the depth information of the whole surface can be obtained.

The overview of the TESP-3D imaging reconstruction procedure is shown in Fig. 3, including two steps. In the first step, the data streams acquired by four SPDs are shown in Fig. 3(a). Herein, a wooden stick (radius 5mm, length 30mm) is fixed on an optical chopper (Thorlabs MC1F10) as the test object. The maximum rotating speed of this chopper is 3000 r/min corresponding to a rotation period of 20 ms. The time for capturing the whole target image is approximately 4 ms. In this system, one dimensional (1D) light patterns are focused on the object. With the chopper rotating, the data stream containing the whole object spectrum is acquired by each SPD. Through the IFFT process four images with different shadow are reconstructed, as shown in Fig. 3(b). In the second step, utilizing the shading information and the detection directions of each SPD, both the surface orientation and the 3D information of the object can be obtained, which is shown in Fig. 3(c).

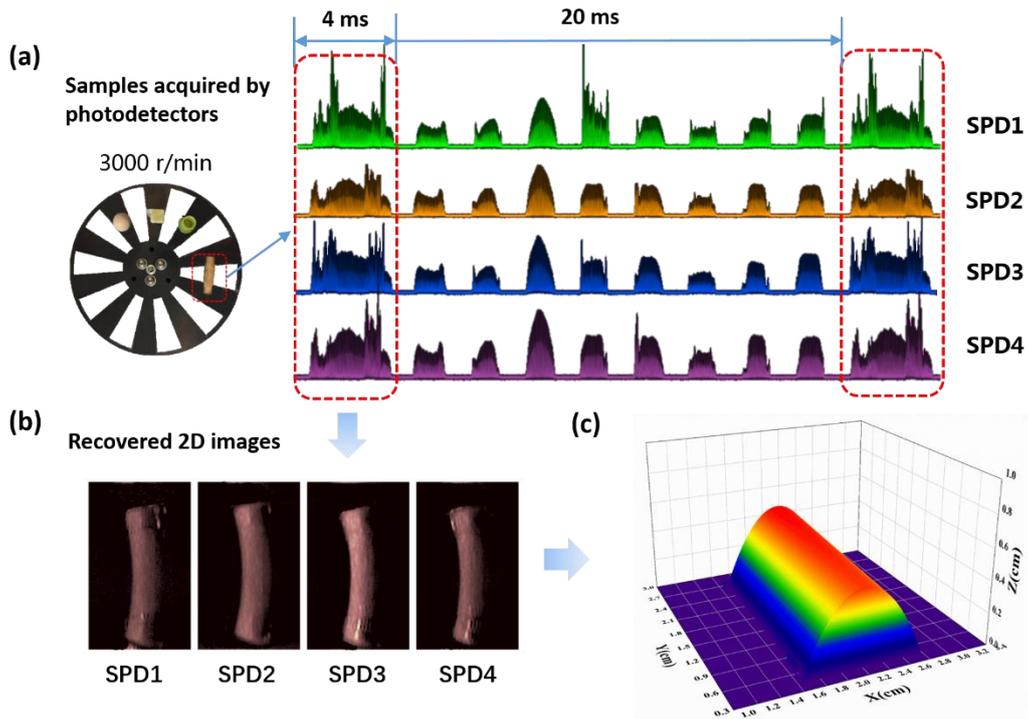

**Figure 3. 3D image reconstruction.** (a) Waveforms of 4 SPDs captured by the oscilloscope. The signals correspond the objects and chopper wheel structure. (b) Recovered 2D images of the wooden stick from 4 points of views with different shadow information. (c) Reconstructed 3D image of the object.

To test the system performance, we put different samples (a hemisphere, a declining plane, a wheel and a column) on the rotating chopper with rotating speed of 3000 r/min (corresponding to the linear speed of 25 m/s), and the structured light lines scan the whole geometry of the objects. A 5-GS/s digital oscilloscope is applied for data acquisition. The photometric stereo algorithm in this demonstration is realized in a personal computer, and this algorithm can be easily transplanted into a GPU hardware for speeding up the computation. Figure 4 shows the 3D shape profile of the half chopper within 10 ms, including all the samples and the blades. More fine temporal resolution images within a 4-µs interval are also presented in Fig.4. Then with the rotating procedure, each reconstructed depth line can be accumulated to form the 3D geometry of the objects. A 500x slow motion video of the line-scanning 3D images is also available in Supplementary Movie 2. As a comparison, we also use an iPhone X in SLomo mode (120 fps) and a fast infrared CCD (Xenics Xeva-1.7-640, 100 fps) to capture the high-speed 3D events, the recorded videos are also shown in the sub-windows of Movie 2.

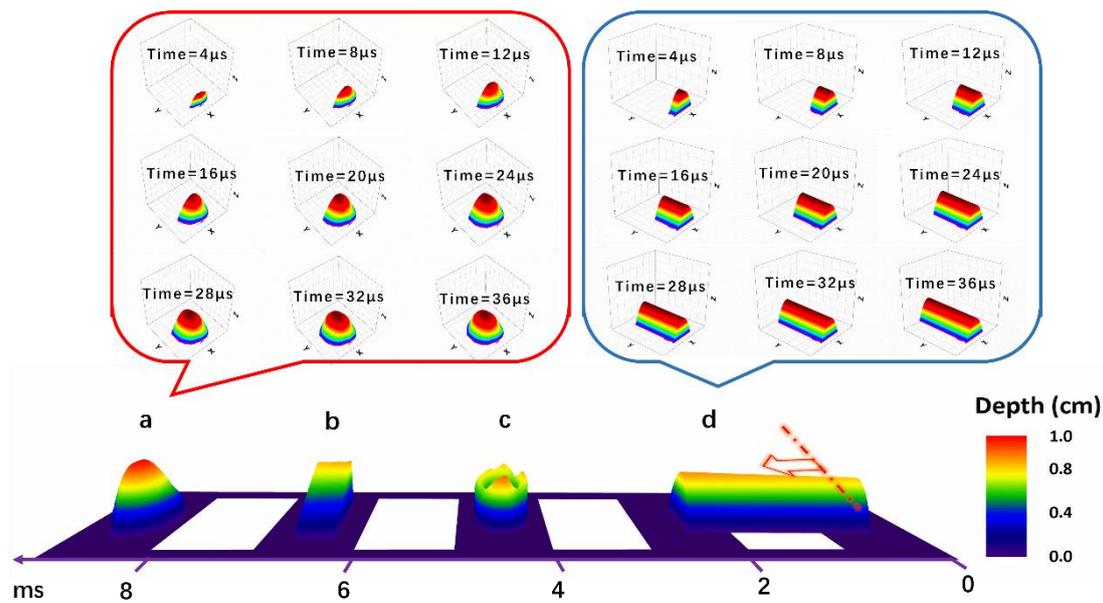

**Figure 4. Real-time 3D imaging.** The scanned imaging of the 3D objects on the high-speed rotating chopper.

Here, we have demonstrated a high-speed 3D imaging system TESP-3D, which is capable of capturing 3D geometry information of fast moving objects. Multi-viewing photodetectors are used to reconstruct images with different shadow in a line frame rate of 500,000 fps. It can be easily expanded to planar imaging by using 2D dispersion devices[21,30]. TESP-3D creatively combines the time-encoded technique and the single-pixel imaging technique to overcome the speed limitations on conventional pattern generation and the sluggish responsivity of CCD and CMOS cameras. The speed of illumination pattern generation in TESP-3D is only determined by the laser pulse repetition rate and it can achieve generation rate of GHz, which is over 1000 times faster than conventional digital micro-mirror devices. So even after thousands or millions of iterations used for single-pixel reconstruction, TESP-3D imaging speed could also reach up to more than 100 kHz. In our demonstration, for the rotation speed limitation of the chopper, the maximum moving speed of the objects is only 25 m/s. However, in principle the speed will be up to 500 m/s with one millimeter resolution. TESP-3D opens a gate for capturing 3D images in micro-seconds for the first time. It's a promising solution for high throughput industrial on-line inspection or autopilot screening in the future information society.

**Acknowledgements**

We thank Prof. Guijin Wang from Tsinghua University for useful discussions. This work is supported by NSFC under Contract 61771284, Beijing Natural Science Foundation under Contracts L182043 and Beijing Municipal Science & Technology Commission with Grant No. Z181100008918011.

**Author contribution statement**

J. Teng, Q. Guo and H. Chen conceived the high-speed 3D single-pixel imaging scheme and designed the experimental system. J. Teng and Q. Guo performed the experiment, collected the data, and processed the data. J. Teng, Q. Guo and H. Chen wrote and revised the manuscript. H. Chen supervised the project. All the authors reviewed this manuscript.

**Supplementary**

**Optical source for TESP-3D system.** In the TESP-3D system, the light source is a mode-locked laser (MLL) emitted the laser pulses in the repetition rate of 50 MHz, and the center wavelength is 1560nm with the pulse width is 150 fs. When the optical pulses go through the dispersion compensating fiber (DCF), whose group velocity dispersion (GVD) is -1368 ps/nm. The pulse width expands in the time domain to 20ns.

**Time-encoded 1D illumination pattern generating procedure**. Utilizing the Mach-Zehnder modulator (MZM) to modulate the sinusoidal electric signal generated by an arbitrary waveform generator (AWG) (Tektronix AWG7000A) synchronized with the MLL. Via the 12.5-Gb/s MZM (PHOTLINE MA-LN-10) the sinusoidal signal with mutative frequency and phase is mapped to the optical pulse. With the pulse going through the single mode fiber (SMF) with dispersion 1368 ps/nm, the time-stretch pulses are compressed to the narrow pulses again, but the pulse spectra are modulated same with the designed sinusoidal pattern in the AWG. Subsequently, a fiber port collimator (THORLABS PAFA-X-4-C) is used to produce a collimated beam incident on a diffraction grating (DG) whose groove density is 1200/mm. By the means of the angular dispersion of the diffraction grating, a 1D structured illumination pattern is generated, which is same as the signal designed in the AWG.

**Single-pixel imaging via Fourier spectrum acquisition.** A phase-shifting technique is used to perform Fourier spectrum acquisition, and the patterns are shown in the Fig.5. In our system, there is an illumination pattern with the unique frequency $f_{rep}$ and phase $\varphi_j$ (0, π / 2, π or 3π/2) projecting on the object. The left figure in Fig.5 shows the intensity distribution curve of the 1D structured illumination patterns, and the right figure shows the corresponding spatial patterns. Sinusoidal signals with a specific frequency $f_i = i \cdot f_{rep}$ and four initial phases $\varphi_j$ ($0$, $\pi/2$, $\pi$ and $3\pi/2$) are used to modulate the intensities of every four adjacent pulses.(the whole procedure is shown in the Fig.1 b) The output signal of the MZM $P_{out}(t, f_i)$ is given by:

$$P_{out}(t, f_i) = \frac{1}{2} P_{in}(t)[1 + \alpha \cos(2\pi f_i t + \varphi)] \tag{1}$$

where $P_{in}(t)$ denotes the input optical pulse train. $\alpha = \pi / V_\pi$ is the modulation index of the MZM, $V_\pi$ is its half-wave voltage. The spectrally encoded optical pulses are compressed, collimated into free space and then incident onto a diffraction grating. Through a linear wavelength-to-space mapping, 1D sinusoidal light patterns are generated, which can be written as:

$$C_{out}(x, f_x) = A + B \cos(2\pi f_x x + \varphi) \tag{2}$$

where A and B account for the average brightness and the fringe contrast, respectively. $f_x$ is the spatial frequency corresponding to $f_i$ resulting from the linear time-to-wavelength-to-space mapping, where $x$ denotes the Cartesian coordinate in 1D space. After illuminating a scene, the backscattered light is recorded by a high-speed photodiode. The acquired measurements are expressed as:

$$V_\varphi(f_x) = K \int_x I(x) C_{out}(x, f_x) dx + V_n \tag{3}$$

where $I(x)$ is the scene reflectivity. $K$ is a constant and $V_n$ is the noise term. Each complex Fourier coefficient can be obtained from four measurements ($V_0$, $V_{\pi/2}$, $V_\pi$ and $V_{3\pi/2}$) as follows:

$$[V_0(f_x) - V_\pi(f_x)] + j \cdot [V_{\pi/2}(f_x) - V_{3\pi/2}(f_x)] = K \omega_{f_x} \tag{4}$$

where $\omega_{f_x}$ is the coefficient of the spatial frequency $f_x$. When $M$ low-frequency coefficients (namely $[\omega_0, \omega_{f_0}, \omega_{2f_0}, \cdots, \omega_{Mf_0}]$ ) are acquired and the high-frequency coefficients (namely $[\omega_{(M+1)f_0}, \omega_{(M+2)f_0}, \cdots, \omega_{Nf_0}]$) are all set to zero. Inverse discrete Fourier transform (IDFT) can be performed to reconstruct $I(x)$ as follows:

$$\begin{bmatrix} I(0) \\ I(1) \\ M \\ I(M) \\ I(M+1) \\ M \\ I(N) \end{bmatrix} = \frac{1}{N} \begin{bmatrix} 1 & 1 & L & 1 & 1 & L & 1 \\ 1 & W_N^{-1} & L & W_N^{-M} & W_N^{-(M+1)} & L & W_N^{-(N-1)} \\ M & M & O & M & M & O & M \\ 1 & W_N^{-M} & L & W_N^{-M \cdot M} & W_N^{-M(M+1)} & L & W_N^{-M(N-1)} \\ 1 & W_N^{-(M+1)} & L & W_N^{-(M+1)M} & W_N^{-(M+1)(M+1)} & L & W_N^{-(M+1)(N-1)} \\ M & M & O & M & M & O & M \\ 1 & W_N^{-(N-1)} & L & W_N^{-(N-1)M} & W_N^{-(N-1)(M+1)} & L & W_N^{-(N-1)(N-1)} \end{bmatrix} \begin{bmatrix} \omega_0 \\ \omega_{f_0} \\ M \\ \omega_{(M-1)f_0} \\ 0 \\ M \\ 0 \end{bmatrix} \tag{5}$$

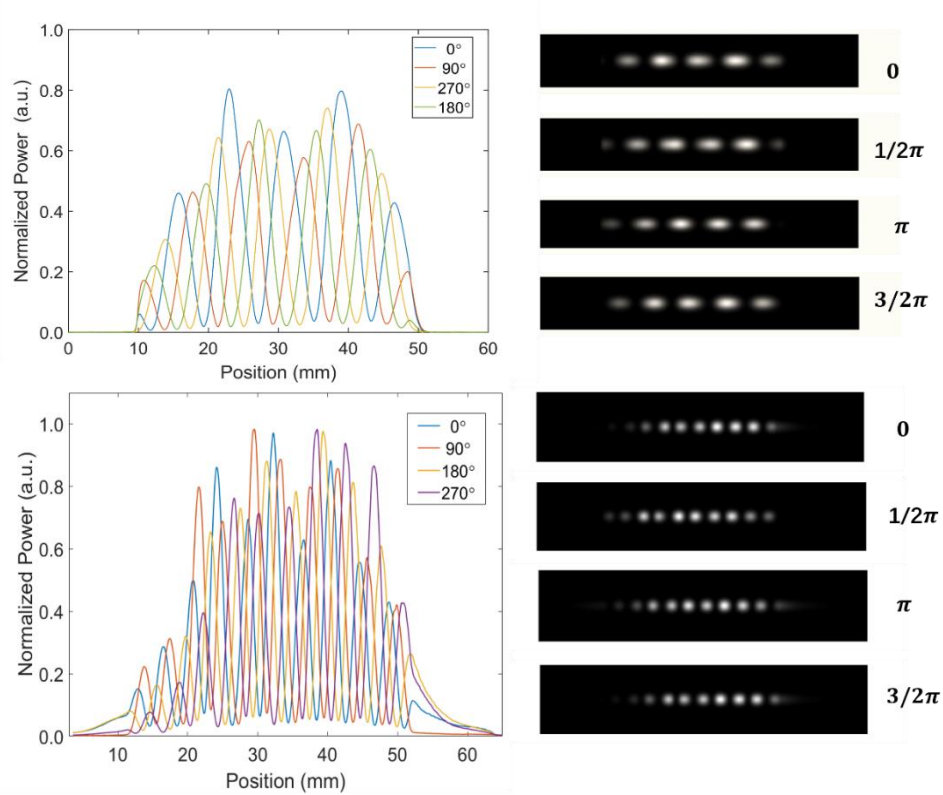

**Fig.5** The four phase-shifting illumination patterns generated by the TEP.

**3D line-scanning imaging based on photometric stereo**. In our system, the object fixed on the rotating platform, optical chopper (Thorlabs MC1F10), is moving with the line speed of 25 m/s. And there are four single-pixel photo detectors (Throlabs DET10N2) located on four different positions to detecting the backscattering illumination, which is encoded by the phase-shafting sinusoidal pattern. With the fast sampling process of four SPDs, the raw data will be used to reconstruct a series of 1D images in the time order. During the rotating time period (typically 20 ms), it is able to reconstruct tens of thousands 1D images (each 1D frame needs 2 us). Then accumulating these line images, the 2D image with the object is revealed. Utilizing the four known viewing directions and the corresponding 2D images, it is able to reconstruct the 3D information of the object. Owing to the reconstruction is starting from the center and working outside, the height of the surface at a given point is estimated from nearest-neighbor point utilizing the gradient and the height of the point. The estimation of the surface geometry is performed to get the relatively depth information among all the pixels, so the further work is find a suitable scale with the reference plane to recover the real 3D object.

**The 3D reconstruction effect of all objects and their error map.** The objects on the chopper and their 3D reconstruction effect are shown in the Fig.6. It can find that for the simple objects like declining plane and hemisphere, the reconstruction error (less than 0.2mm) is relatively small compare with the complicated objects with discontinuous surface, like the wheel (0.8mm). The edge part appears larger reconstruction error, which can be decreased with more coded illumination patterns projecting on the surface or increasing the number of the viewing detectors. What's more, in the experiment the material of the surface may also influence the imaging effect, especially for some objects made by transparent or semitransparent material , which will weaken the backscattered signal detecting by the single-pixel camera and yield low SNR (signal noise ratio) in the edge of the objects.

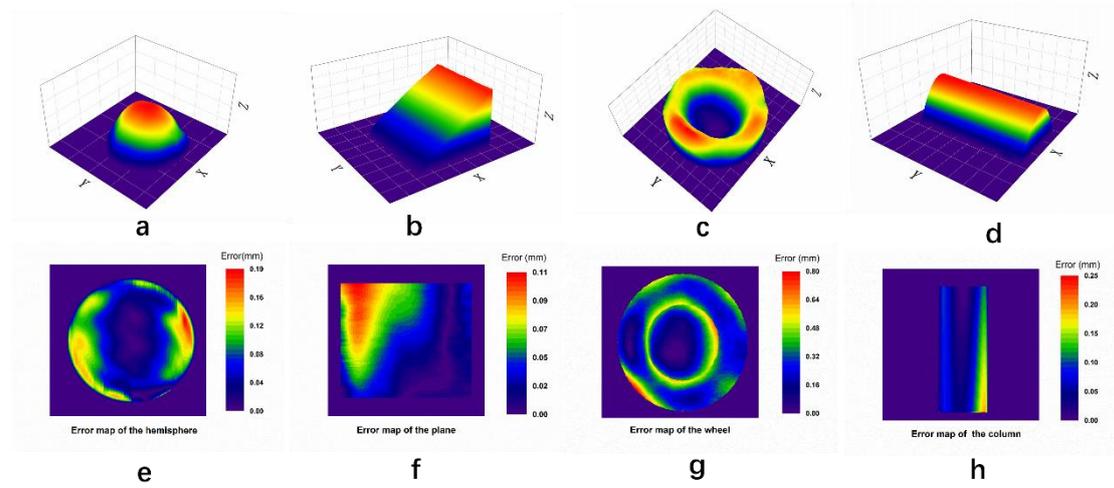

**Fig.6** The reconstruction of four 3D objects (plane, sunken wheel, column and hemisphere) & the error Map of these objects